\documentclass[sigconf,nonacm]{acmart}
\AtBeginDocument{%
  }

\usepackage{longtable}   
\usepackage{booktabs}    
\usepackage{array}       
\usepackage{tabularx}  
\usepackage{ragged2e}  
\usepackage[table]{xcolor}
\usepackage{subcaption}

\setcopyright{acmlicensed}
\copyrightyear{2018}
\acmYear{2018}
\acmDOI{XXXXXXX.XXXXXXX}
\acmConference[Conference acronym 'XX]{Make sure to enter the correct
  conference title from your rights confirmation email}{June 03--05,
  2018}{Woodstock, NY}
\acmISBN{}

\begin{document}

\title{Between Knowledge and Care: Evaluating Generative AI-Based IUI in Type 2 Diabetes Management Through Patient and Physician Perspectives}

\author{Yibo Meng}
\authornote{Both authors contributed equally to this research.}
\affiliation{%
  \institution{Tsinghua University}
  \city{Beijing}
  \country{China}
}
\email{mengyb22@mails.tsinghua.edu.cn}

\author{Ruiqi Chen}
\authornotemark[1]
\affiliation{%
  \institution{University of Washington}
  \city{Seattle}
  \state{Washington}
  \country{United States}
}
\email{ruiqich@uw.edu}

\author{Bingyi Liu}
\affiliation{%
  \institution{University of Michigan, Ann Arbor}
  \city{Ann Arbor}
  \state{Michigan}
  \country{United State}
}
\email{bingyi@umich.edu}

\author{Yan Guan}
\affiliation{%
  \institution{Tsinghua University}
  \city{Beijing}
  \country{China}
}
\email{guany@tsinghua.edu.cn}

\author{Xiaolan Ding}
\affiliation{%
  \institution{North China University of Science and Technology Health Science Center}
  \city{Tangshan}
  \country{China}
}

\renewcommand{\shortauthors}{Meng et al.}

\begin{abstract}
  Generative AI systems are increasingly used by patients seeking everyday health guidance, yet their appropriateness in chronic care contexts remains unclear. Focusing on Type~2 Diabetes Mellitus (T2DM), this paper presents a mixed-methods investigation into how AI-generated health information is interpreted by patients and evaluated by physicians in China. Drawing on formative patient grounding and a dimension-based physician evaluation, we examine AI responses along five quality dimensions: \textit{Accuracy}, \textit{Safety}, \textit{Clarity}, \textit{Integrity}, and \textit{Action Orientation}. Our findings reveal that while current systems perform well in factual explanation and general lifestyle guidance, they frequently break down in safety signaling, contextual judgment, and responsibility boundaries—particularly when fluent responses invite overtrust. By treating quality dimensions as an interpretive lens rather than a fixed framework, this work highlights the need for intelligent user interfaces that actively mediate AI outputs in chronic disease management, supporting calibrated trust and responsible boundary-setting in long-term care.
\end{abstract}



\maketitle

\section{INTRODUCTION}
Generative AI systems are rapidly entering consumer-facing health contexts, raising renewed interest in how large language models (LLMs) might support chronic disease self-management. While recent work demonstrates that LLMs can synthesize biomedical knowledge and produce fluent explanations \cite{samimi2025visual,hernandez2023future}, fluency and coverage alone do not guarantee that health information is safe, actionable, or interactionally appropriate for real users \cite{akrimi2025chatgpt}. These concerns are particularly salient in chronic care, where patients repeatedly seek guidance under uncertainty, and where small omissions or poorly framed advice can accumulate into meaningful risk over time \cite{10.1145/3544548.3581251,mirbabaie2025digital,mitchell2021automated}.

In this work, we focus on Type~2 Diabetes Mellitus (T2DM) as a representative chronic condition to examine the role boundaries and reliability of generative AI in everyday self-management. T2DM requires sustained decisions around medication, diet, physical activity, and psychosocial regulation \cite{gong2020my}. In China, where clinical workloads are high and specialist access is uneven, patients increasingly turn to always-available digital sources---including generative AI tools---to interpret symptoms, translate medical terminology, and explore self-care options outside the clinic \cite{hernandez2023future,lanshan2025factors,guo2025promoting}. This usage creates a tension central to HCI and IUI: AI can lower barriers to health information access, yet it can also invite overreliance when its outputs appear confident but fail to reflect contextual constraints and responsibility boundaries.

A core limitation in prior evaluations is that health information quality is often operationalized as a largely one-dimensional property---typically correctness on benchmark-style question answering---despite mounting evidence that patients and clinicians attend to additional dimensions such as safety signaling, coherence, and the practical framing of next steps \cite{hernandez2023future,akrimi2025chatgpt}. More broadly, HCI research has shown that interface fluency can inflate perceived credibility, complicating calibrated trust in safety-critical settings \cite{10.1145/3544548.3581251}. What remains underexplored is how clinicians interpret these quality dimensions in practice when reading AI-generated responses to authentic patient questions, and how such dimension-level judgments can inform the design of health IUIs that support appropriate reliance rather than superficial trust.

To address this gap, we conducted a two-phase mixed-methods study. Phase~1 served as a formative grounding study to capture how people with T2DM engage with generative AI in real-world self-management contexts and to inform ecologically grounded evaluation stimuli. Phase~2 then conducted a physician-centered evaluation of four widely used generative AI systems using a dimension-based rubric that reflects clinically meaningful criteria for information adequacy and risk. Across quantitative ratings and follow-up interviews, we examine how models exhibit systematic trade-offs across \textit{accuracy}, \textit{safety}, \textit{clarity}, \textit{integrity}, and \textit{action orientation}, and how these trade-offs shape clinicians' trust judgments.

Our contributions are threefold: (1) a dimension-based expert evaluation of contemporary generative AI systems on patient-authored T2DM questions, highlighting systematic trade-offs rather than a single ``best model'' narrative; (2) qualitative evidence on how physicians operationalize quality dimensions through concrete judgment moments and failure modes; and (3) design implications for dimension-aware health IUIs, including mediation strategies that calibrate trust by aligning actionability and presentation with clinically grounded risk boundaries. Together, our findings argue that chronic care contexts demand a shift from answer-centric conversational agents toward IUIs that actively mediate model outputs to support safe, context-sensitive reliance.

\section{RELATED WORK}
\subsection{Information Needs and Quality Judgments in AI-Mediated Chronic Care}

Type~2 diabetes mellitus is a chronic condition that requires patients to engage in sustained information seeking and self-management across medication, diet, physical activity, and emotional regulation \cite{gong2020my, gerstenberg2025living}. Prior work in HCI and health informatics has shown that patients often struggle to interpret clinical guidelines, contextualize numerical indicators such as HbA1c, and translate abstract medical knowledge into everyday action \cite{hernandez2023future, mayberry2019mhealth}.

In practice, patients frequently supplement limited clinical consultations with online sources and AI-powered tools, particularly for routine questions and early-stage sensemaking \cite{10.1145/3290605.3300600, akrimi2025chatgpt}. Importantly, studies consistently show that patient needs extend beyond factual explanation to include lifestyle adaptation, behavioral guidance, and emotional coping \cite{biernatzki2018information, dsouza2024identification}. These findings suggest that the perceived usefulness of health information is shaped not only by correctness, but also by how well information is contextualized, interpreted, and translated into action.

Within HCI, T2DM has served as a recurring design context for intelligent health technologies such as education tools, self-tracking systems, and coaching interfaces \cite{ayobi2021co, bhattacharya2023directive}. While these systems demonstrate the value of tailored feedback, they typically rely on rule-based logic or pre-scripted content. The emergence of generative AI introduces new flexibility in health information access, but also raises questions about how patients assess the appropriateness, safety, and actionability of AI-generated content in chronic care contexts.

\subsection{Evaluating and Mediating AI-Generated Health Information}

Recent advances in large language models (LLMs) have enabled conversational access to health information on topics such as diet, exercise, and medication management \cite{reddy2024generative, rebitschek2025evaluating}. At the same time, growing evidence suggests that AI-generated health content may be inaccurate, incomplete, or misleading when applied to domain-specific scenarios such as diabetes care \cite{guo2025promoting, agarwal2024medhalu}.

Most existing evaluations rely on benchmark-style medical question–answering datasets that prioritize factual correctness at the answer level \cite{jin2019pubmedqa, kim2024medexqa}. While valuable for model comparison, such approaches offer limited insight into how information is interpreted and acted upon by patients. In response, researchers in HCI and intelligent user interfaces have argued for multidimensional evaluation frameworks that account for safety, clarity, and actionability in addition to accuracy \cite{10.1145/3544548.3581251, kim2024human}.

Despite this shift, chronic disease contexts remain underexplored from a clinical evaluation perspective. Existing studies on AI use in T2DM primarily focus on usability or patient satisfaction, with limited involvement of clinicians in assessing risk, appropriateness, and responsibility boundaries \cite{swallow2025digibete, tseng2025designing}. Prior work in health-oriented IUI design highlights the importance of risk-sensitive interaction strategies such as uncertainty communication and deferral to clinicians \cite{mirbabaie2025digital, schombs2025designing}, yet these mechanisms are rarely grounded in systematic clinical judgment.

Together, these gaps motivate the need for clinician-informed, dimension-based evaluation of AI-generated health information, as well as interface-level mediation strategies that support calibrated trust and appropriate use in chronic care settings.

\section{METHOD}
Our study followed a two-phase mixed-methods design. Phase~1 served as a formative grounding study to surface how patients conceptualize and evaluate AI-generated health information in everyday self-management contexts. Phase~2 then built on this grounding to conduct a dimension-based expert evaluation with physicians. Importantly, Phase~1 was not intended to produce standalone empirical findings; instead, it established the evaluative dimensions, stimuli,
and scope for the physician study that follows.

\subsection{Phase 1: Formative Grounding of Evaluation Dimensions}

Phase~1 explored how patients with Type~2 Diabetes Mellitus (T2DM)
interact with generative AI tools in real-world self-management scenarios. The goal of this phase was epistemic rather than confirmatory: to surface the implicit criteria patients use when judging AI-generated health information, and to ground the construction of evaluation dimensions used in Phase~2.

We focused on individuals with type~2 diabetes mellitus, a chronic condition that requires sustained information seeking, behavioral regulation, and interpretive judgment. This context allowed us to observe how patients engage with AI-generated health information across routine, ambiguous, and emotionally laden situations, thereby revealing tensions around correctness, safety,
interpretability, coherence, and actionability that are difficult to capture through expert-authored prompts alone.

\subsubsection{Participants}

We recruited participants through a combination of online and offline channels. Online recruitment notices were posted on widely used Chinese social platforms, including Xiaohongshu, Bilibili, and Baidu Tieba. Offline recruitment was conducted in community centers and public spaces across four provinces in Northern China (Henan, Hebei, Shanxi, and Shandong). Eligibility criteria required participants to be at least 18 years old, have a clinical diagnosis of T2DM, and possess prior experience using generative AI tools (e.g., ChatGPT, Wenxin Yiyan, or similar systems) for diabetes-related information seeking or self-management.

Participants were excluded if they had not been diagnosed with T2DM, were experiencing severe complications or other serious illnesses that could interfere with participation, or had recently taken part in other clinical studies with overlapping objectives. All participants provided informed consent prior to participation.

A total of 21 participants were enrolled, including 11 males and 10 females, ranging in age from 33 to 65 years ($M = 47.76$, $SD = 8.14$). Duration of T2DM ranged from 1 to 23 years ($M = 8.10$, $SD = 5.52$). Twelve participants resided in rural areas and nine in urban settings. Educational backgrounds varied substantially, ranging from no formal education to bachelor’s degrees. Table~\ref{tab:obs_participants} summarizes participant demographics.

The study protocol was reviewed and approved by the Institutional Review Board (IRB) of [Anonymous University]. All collected data were anonymized prior to analysis, and participants retained the right to withdraw or request data deletion at any time. Each participant received a compensation of 30~RMB.

\begin{table*}[htbp]
\centering
\caption{Demographic characteristics of participants in the observational study (N=21).}
\label{tab:obs_participants}
\begin{tabular}{cccccc}
\toprule
ID & Age & Years with T2DM & Gender & Residence & Education \\
\midrule
1  & 44 & 5  & F & Urban & High school \\
2  & 45 & 5  & M & Urban & High school \\
3  & 54 & 4  & M & Rural & Middle school \\
4  & 55 & 12 & M & Urban & Middle school \\
5  & 56 & 13 & F & Rural & Primary school \\
6  & 65 & 23 & F & Urban & Primary school \\
7  & 57 & 17 & M & Rural & None \\
8  & 56 & 11 & F & Urban & High school \\
9  & 45 & 11 & M & Rural & High school \\
10 & 38 & 3  & F & Urban & Bachelor \\
11 & 36 & 3  & M & Urban & Bachelor \\
12 & 39 & 2  & F & Urban & High school \\
13 & 33 & 4  & M & Rural & Bachelor \\
14 & 45 & 11 & M & Rural & High school \\
15 & 56 & 12 & F & Urban & None \\
16 & 52 & 10 & F & Rural & None \\
17 & 49 & 6  & M & Rural & High school \\
18 & 41 & 1  & M & Rural & Bachelor \\
19 & 46 & 3  & F & Urban & Middle school \\
20 & 44 & 7  & M & Rural & None \\
21 & 47 & 7  & F & Rural & None \\
\bottomrule
\end{tabular}
\end{table*}

\subsubsection{Procedure}

Procedures in Phase~1 were intentionally designed to elicit implicit evaluative concerns that patients bring to AI-mediated health interactions, rather than to benchmark usability or system performance. The study followed a multi-stage protocol combining structured questionnaires and semi-structured interviews.

Participants first completed two structured instruments: an AI Attitude Questionnaire and an AI Usage Questionnaire. The attitude questionnaire comprised ten items rated on a 10-point Likert scale (1~=~strongly disagree, 10~=~strongly agree), probing perceptions related to convenience, perceived usefulness, personalization, psychological support, risk awareness, and overall satisfaction. The usage questionnaire captured participants’ prior experiences with generative AI systems, including platforms used, interaction modalities (e.g., mobile applications or embedded assistants), and examples of typical information-seeking scenarios.

To contextualize questionnaire responses and surface evaluative reasoning not easily expressed through fixed-response items, we conducted 35--55 minute semi-structured interviews with each participant. Interview questions explored participants’ medical background, daily self-management practices, experiences
with AI-generated health information, perceived benefits and shortcomings, and expectations for future AI support. Interviews were conducted either in person or via phone, depending on participant availability. With participant consent, all interviews were audio-recorded and transcribed verbatim within 24 hours.
Each transcript was cross-checked to ensure accuracy and completeness.

All research staff received standardized training in qualitative interviewing and completed preparatory reviews on T2DM management to ensure ethically and contextually appropriate data collection. Data were stored securely on the institutional cloud infrastructure of [Anonymous University], with access restricted to authorized researchers.

\subsubsection{Data Analysis}
We adopted a mixed-methods analytic approach that combined descriptive analysis of structured questionnaire data with thematic analysis of qualitative materials. Questionnaire responses were summarized using descriptive statistics to characterize participant backgrounds and general AI usage patterns.

Qualitative data—including interview transcripts and patient-authored
AI queries—were analyzed using thematic analysis \cite{braun2006using}. Two researchers independently coded an initial subset of transcripts to construct a shared codebook, which was iteratively refined through discussion. Inter-coder agreement exceeded 85\%, with disagreements resolved by consensus. Coding combined inductive strategies, which captured emergent patterns in
patients’ information-seeking behaviors, and deductive strategies,
which were informed by preliminary evaluative concerns relevant to
subsequent expert assessment.

Importantly, analytic codes and themes identified in Phase~1 were not
treated as empirical findings to be interpreted or generalized.
Instead, they functioned as sensitizing constructs that informed:
(1) the definition of evaluation dimensions,
(2) the scope of physician judgment criteria, and
(3) the selection of patient-authored questions used as evaluation
stimuli in Phase~2.
By design, insights from Phase~1 are reflected in the structure of the expert evaluation rather than reported as standalone results.

\subsection{Phase 2: Dimension-based Physician Evaluation}
Building on insights from the formative study in Phase~1, we conducted a physician-centered evaluation to assess the quality of AI-generated health information for T2DM management. This phase shifts the analytic perspective from patient experiences to expert clinical judgment, examining how generative AI responses perform along core quality dimensions, including \textit{accuracy}, \textit{clarity}, \textit{safety}, \textit{integrity}, and \textit{action orientation}. The goal of this phase is to identify system-level strengths and limitations as perceived by clinicians, and to inform the design of safer and more trustworthy AI-assisted health interfaces.

\subsubsection{Participants}

We recruited participating physicians through online outreach on major Chinese platforms, including Xiaohongshu, Bilibili, and Baidu Tieba. To be eligible, physicians were required to meet the following criteria: (1) hold a valid medical license; (2) be currently practicing in endocrinology or have long-term experience treating T2DM; (3) have a minimum of five years of clinical experience in T2DM care; and (4) hold the professional rank of attending physician or higher. All participants were required to provide written informed consent. Physicians were excluded if they were directly involved in the development of AI tools for diabetes care or held financial interests in related companies, to avoid conflicts of interest. Physicians currently suspended or not actively practicing were also excluded.

A total of seven physicians participated in this evaluation study. Their ages ranged from 34 to 62 years ($M=48.86$, $SD=11.57$), with years of clinical practice ranging from 5 to 33 years ($M=19.86$, $SD=11.60$). All participants held a PhD degree. Five physicians specialized in endocrinology or metabolic disorders, one in rehabilitation medicine, and one in cardiology. Four practiced in urban settings and three in rural areas. Table~\ref{tab:physicians} summarizes the demographic characteristics of the participating physicians.

This study was approved by the Institutional Review Board (IRB) of [Anonymous University]. All participants were fully informed of the study's goals, procedures, and potential implications. They retained the right to withdraw from the study or request data deletion at any time. All data were anonymized during collection and analysis. Each physician received a compensation of 100 RMB for their participation.

\begin{table*}[htbp]
\centering
\caption{Demographic characteristics of participating physicians (N=7).}
\label{tab:physicians}
\begin{tabular}{ccccccc}
\toprule
ID & Age & Years of Practice & Department & Gender & Residence & Degree \\
\midrule
1 & 57 & 28 & Endocrinology & M & Urban & PhD \\
2 & 62 & 33 & Endocrinology & M & Urban & PhD \\
3 & 57 & 29 & Endocrinology & M & Urban & PhD \\
4 & 55 & 25 & Rehabilitation & F & Rural & PhD \\
5 & 34 & 5  & Endocrinology and Metabolism & M & Urban & PhD \\
6 & 42 & 13 & Cardiology & F & Urban & PhD \\
7 & 35 & 6  & Endocrinology & F & Rural & PhD \\
\bottomrule
\end{tabular}
\end{table*}

\subsubsection{Procedures}
The evaluation procedure consisted of three tightly coupled steps: clinician-guided question curation, independent AI output evaluation, and post-evaluation expert reflection. Together, these steps enabled a systematic examination of AI-generated health information grounded in real patient queries and expert clinical judgment.

\paragraph{Clinician-guided Question Curation.}
To ensure realism and clinical relevance, the physician team first curated a fixed set of patient-authored questions drawn from the corpus collected in Phase~1. Rather than constructing a taxonomy or performing category-based sampling, this step served solely as stimulus preparation for the evaluation task.

Questions were selected through iterative discussion based on three criteria: (1) frequency of occurrence in patient submissions; (2) clinical significance for diabetes self-management; and (3) potential risk if answered incorrectly or acted upon without professional supervision. This process resulted in a representative set of real-world patient questions that balanced common informational needs with clinically sensitive scenarios. The curated question set was used consistently across all evaluated AI systems.

\paragraph{AI Output Evaluation.}
Each curated question was independently entered into four widely used generative AI systems: GPT-4-turbo~\cite{openai_chatgpt_2025}, DeepSeek-R1~\cite{deepseek_2025}, Kimi K2~\cite{kimi_2025}, and ERNIE Bot 4.5~\cite{erniebot_2025}. To ensure comparability, each query was submitted in a fresh session, with memory or context-retention features disabled where possible. All AI-generated responses were saved verbatim and anonymized by model and question ID.

Physicians independently evaluated each AI response using a standardized five-dimensional rubric developed collaboratively by the clinical team. The evaluation dimensions were:

\begin{itemize}
  \item \textbf{Accuracy}: alignment with current clinical knowledge and evidence-based practice;
  \item \textbf{Safety}: appropriate signaling of risks, contraindications, and the need for professional consultation;
  \item \textbf{Clarity}: accessibility of language and transparency of explanation;
  \item \textbf{Integrity}: internal coherence, completeness, and prioritization of information;
  \item \textbf{Action Orientation}: provision of concrete and practically useful guidance without overstepping clinical authority.
\end{itemize}

Each dimension was scored on a fixed numerical scale. Physicians were instructed to base their ratings on clinical judgment rather than personal preferences or model familiarity. In addition to numerical scores, physicians could optionally record brief qualitative comments to highlight perceived strengths, concerns, or safety issues in specific responses.

\paragraph{Post-evaluation Expert Reflection.}
Following the scoring task, we conducted semi-structured interviews with each physician to probe their evaluative reasoning and interpretation of dimension-level trade-offs. The interview protocol focused on how physicians made judgments about safety, actionability, coherence, and trustworthiness, and why certain responses were perceived as appropriate, risky, or misleading.

Rather than prompting discussion around question categories, interviews centered on dimension-level failure modes, boundary conditions for acceptable action-oriented guidance, and the interaction between fluency and clinical reliability. Interviews also explored physicians’ broader expectations for how generative AI should be integrated into clinical workflows and patient-facing systems. Each interview lasted 30–60 minutes, was audio-recorded with consent, transcribed verbatim, and verified for accuracy.

\subsubsection{Data Analysis}

Our analysis focused on dimension-based evaluation of model quality, aligning with the paper's scope of characterizing trade-offs across \textit{Accuracy}, \textit{Safety}, \textit{Clarity}, \textit{Integrity}, and \textit{Action Orientation}. Although the patient-authored question set covered diverse real-world concerns, we did not treat question type as an analytic factor in this short paper; instead, questions served as ecologically grounded stimuli to probe dimension-level reliability.

We first screened physician rating data for completeness and consistency, and aggregated scores at the model $\times$ dimension level. For each AI model, we computed descriptive statistics (mean, standard deviation, and interquartile range) for each of the five dimensions, as well as an overall score obtained by summing dimension scores (when reported). To examine whether models differed in quality, we conducted repeated-measures analyses with \textit{Model} (four levels) and \textit{Dimension} (five levels) as within-subject factors. When sphericity assumptions were violated, Greenhouse--Geisser corrections were applied. We performed post-hoc pairwise comparisons with multiplicity correction and reported effect sizes to characterize practical significance. Visualizations (e.g., boxplots, radar charts, and heatmaps) were used to summarize model performance profiles across dimensions and to highlight stability versus variability in physician judgments.

Interview transcripts were analyzed using reflexive thematic analysis \cite{braun2006using} to capture physicians' reasoning about dimension-level judgments, trade-offs, and failure modes. Two researchers conducted iterative coding across all transcripts, beginning with open coding to identify recurrent evaluative heuristics (e.g., how clinicians detect weak safety signaling or insufficient prioritization), followed by axial organization around the five evaluation dimensions and cross-cutting themes (e.g., fluency-driven overtrust, responsibility boundaries, and conditions for acceptable action-oriented guidance). Disagreements were resolved through discussion, and representative quotes were selected to illustrate how clinicians operationalized quality dimensions in practice and how these interpretations informed design expectations for dimension-aware health IUIs.

\section{RESULTS}
\subsection{Quantitative Findings}

\subsubsection{Overall Performance Variability Across Models}

\begin{figure}[htbp]
    \centering
    \includegraphics[width=0.5\linewidth]{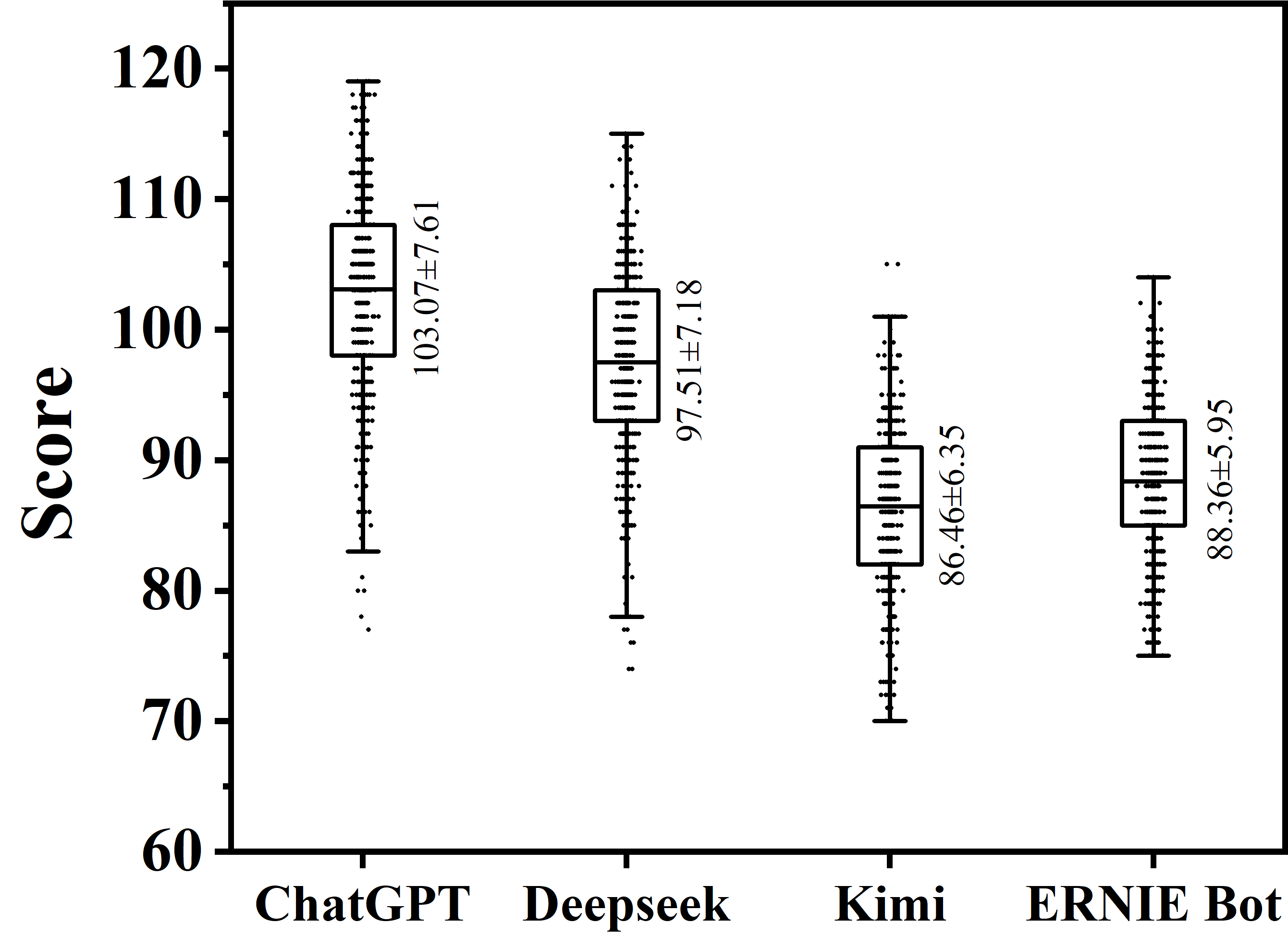}
    \caption{Overall quality evaluation of four AI models. 
    Each box represents the distribution of aggregated scores across all evaluation dimensions and questions for one AI system.}
    \label{fig:overall_quality}
\end{figure}

To assess the overall quality of health information generated by different AI models, we aggregated physician ratings across all questions and evaluation dimensions. As shown in Figure~\ref{fig:overall_quality}, ChatGPT achieved the highest mean score ($M=103.07$, $SD=7.61$), followed by DeepSeek ($M=97.51$, $SD=7.18$). ERNIE Bot ($M=88.36$, $SD=5.95$) and Kimi ($M=86.46$, $SD=6.35$) received considerably lower ratings. In addition to higher means, ChatGPT and DeepSeek displayed narrower interquartile ranges and fewer outliers, indicating more consistent performance. In contrast, Kimi and ERNIE Bot exhibited larger dispersion and more frequent outliers, suggesting unstable behavior across diverse evaluation contexts.

A repeated-measures ANOVA confirmed a significant effect of model type on overall quality, $F(3,18) = 2315.46$, $p < .001$, $\eta^2_G = 0.72$. Post-hoc tests showed that ChatGPT significantly outperformed all three alternatives ($p < .001$ for each), and DeepSeek also scored higher than both Kimi and ERNIE Bot. No reliable difference was observed between the latter two. Together, these results reveal a pronounced performance hierarchy, with ChatGPT providing the most consistent overall output, while the lower-ranked systems showed greater volatility in quality and presentation.

\subsubsection{Dimension-wise Performance Profiles and Trade-offs}

\begin{figure}[htbp]
    \centering
    \begin{subfigure}[t]{0.48\linewidth}
        \centering
        \includegraphics[width=\linewidth]{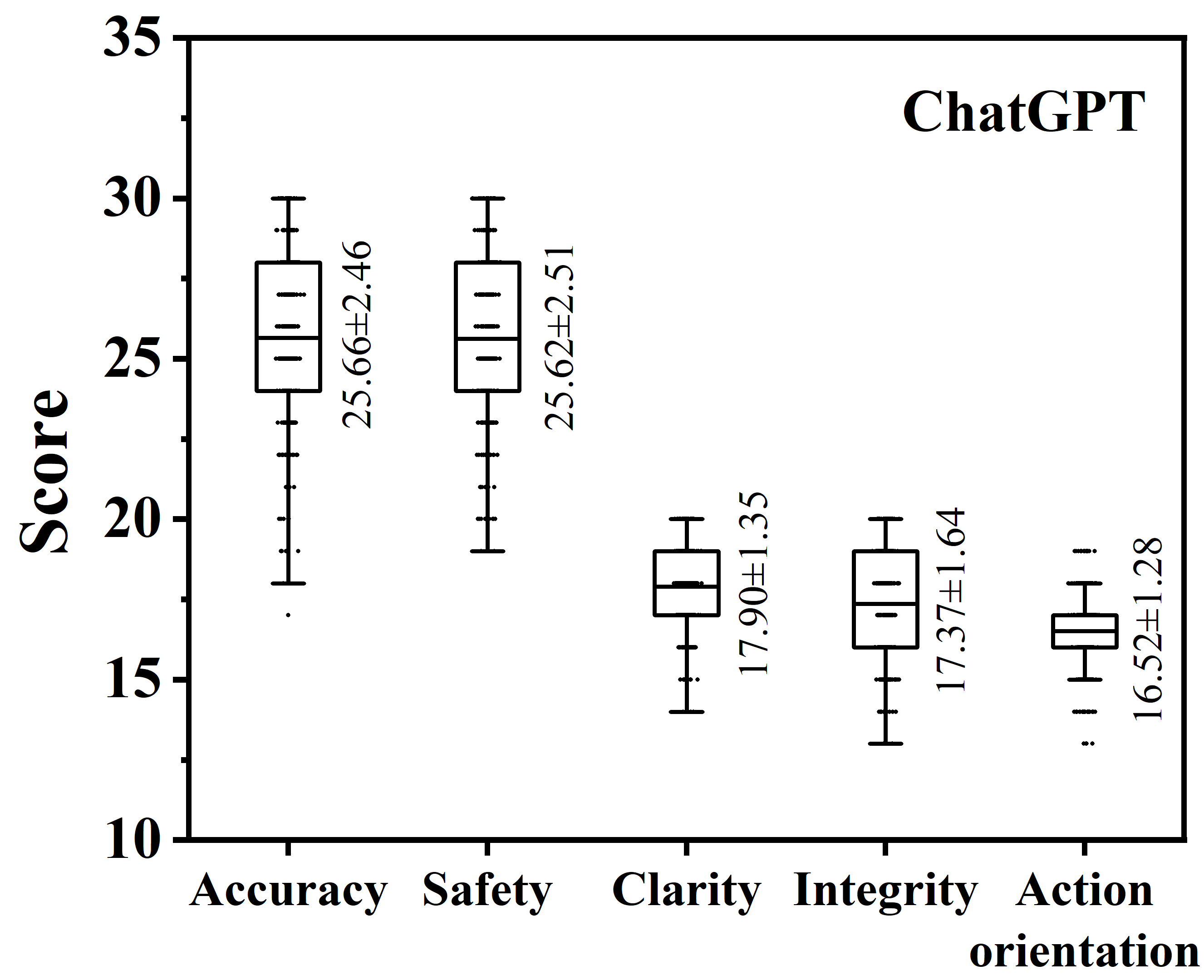}
    \end{subfigure}
    \hfill
    \begin{subfigure}[t]{0.48\linewidth}
        \centering
        \includegraphics[width=\linewidth]{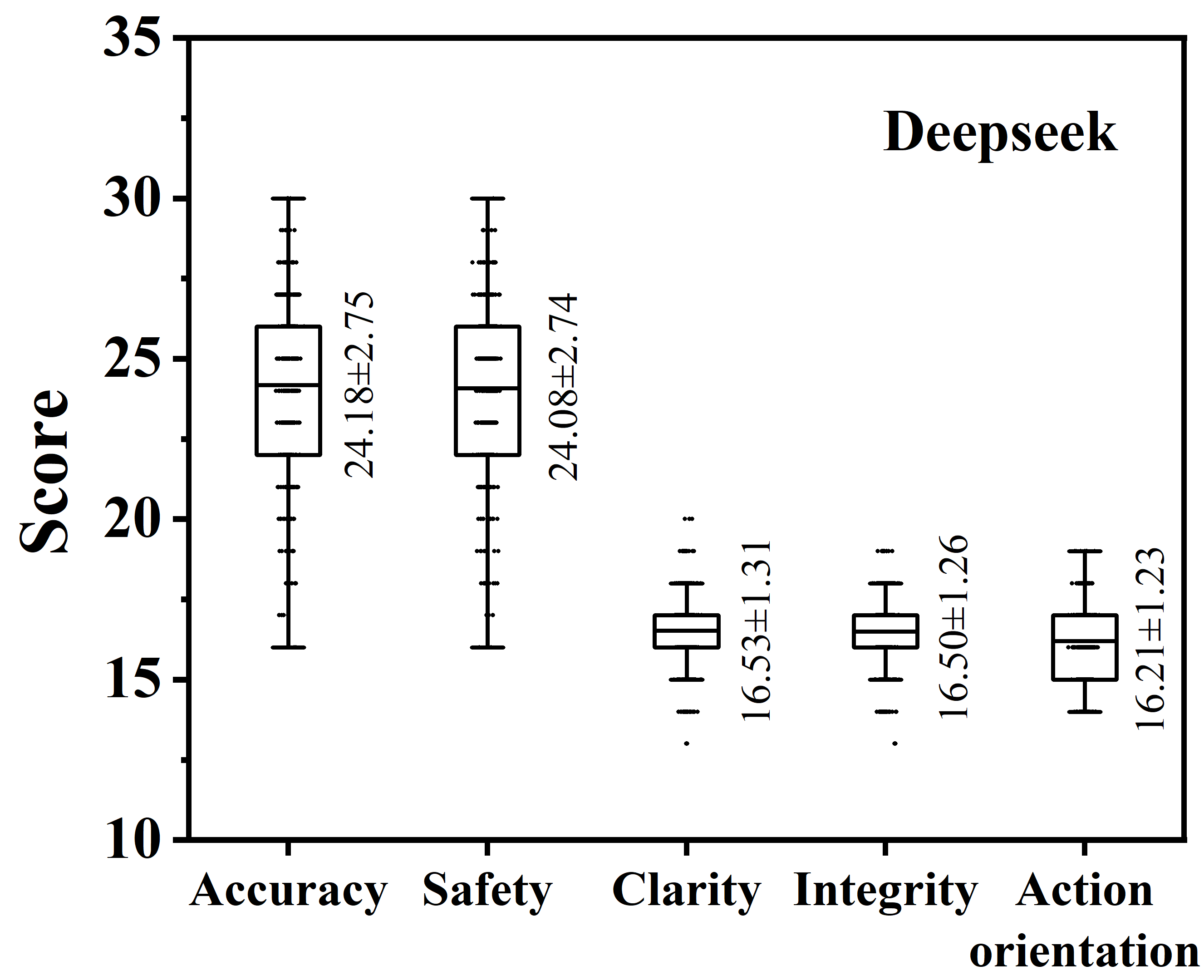}
    \end{subfigure}

    \vspace{1em} 

    \begin{subfigure}[t]{0.48\linewidth}
        \centering
        \includegraphics[width=\linewidth]{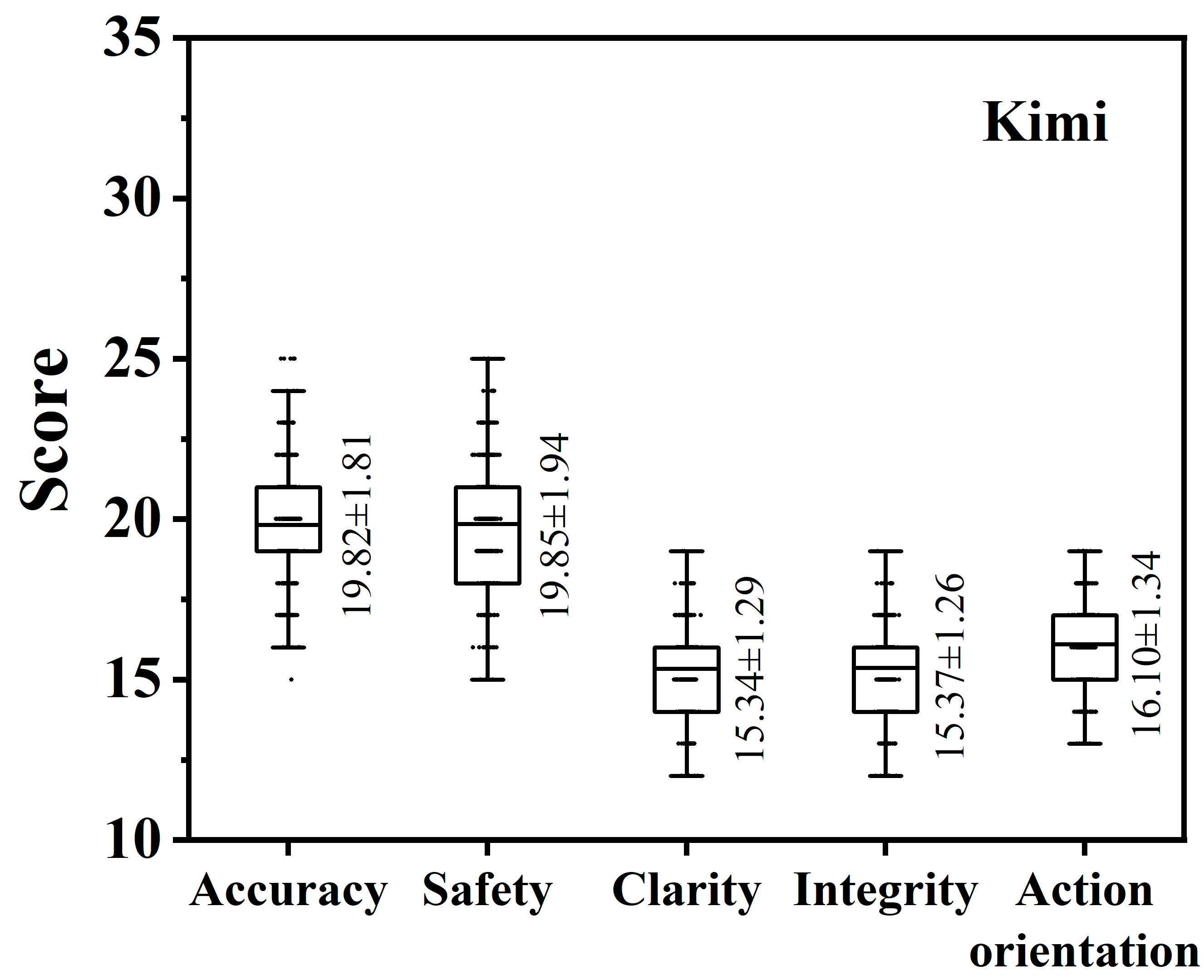}
    \end{subfigure}
    \hfill
    \begin{subfigure}[t]{0.48\linewidth}
        \centering
        \includegraphics[width=\linewidth]{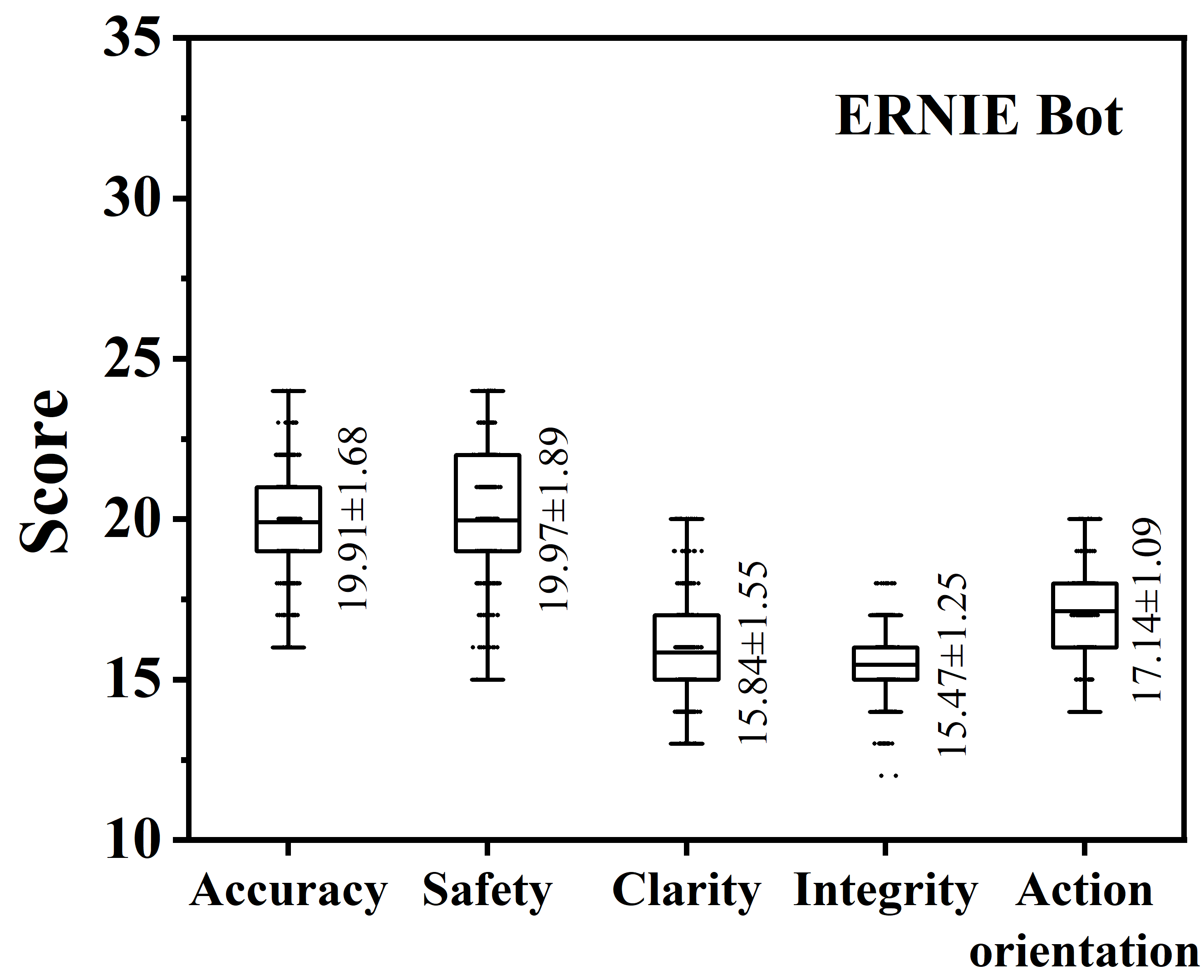}
    \end{subfigure}

    \caption{Dimension-wise quality evaluation of four AI models across five key criteria: Accuracy, Safety, Clarity, Integrity, and Action Orientation. 
    Each subplot represents one AI model. Boxplots depict the distribution of scores across all evaluated questions under each criterion.}
    \label{fig:dimension_wise}
\end{figure}

\begin{figure}[htbp]
    \centering
    \includegraphics[width=0.6\linewidth]{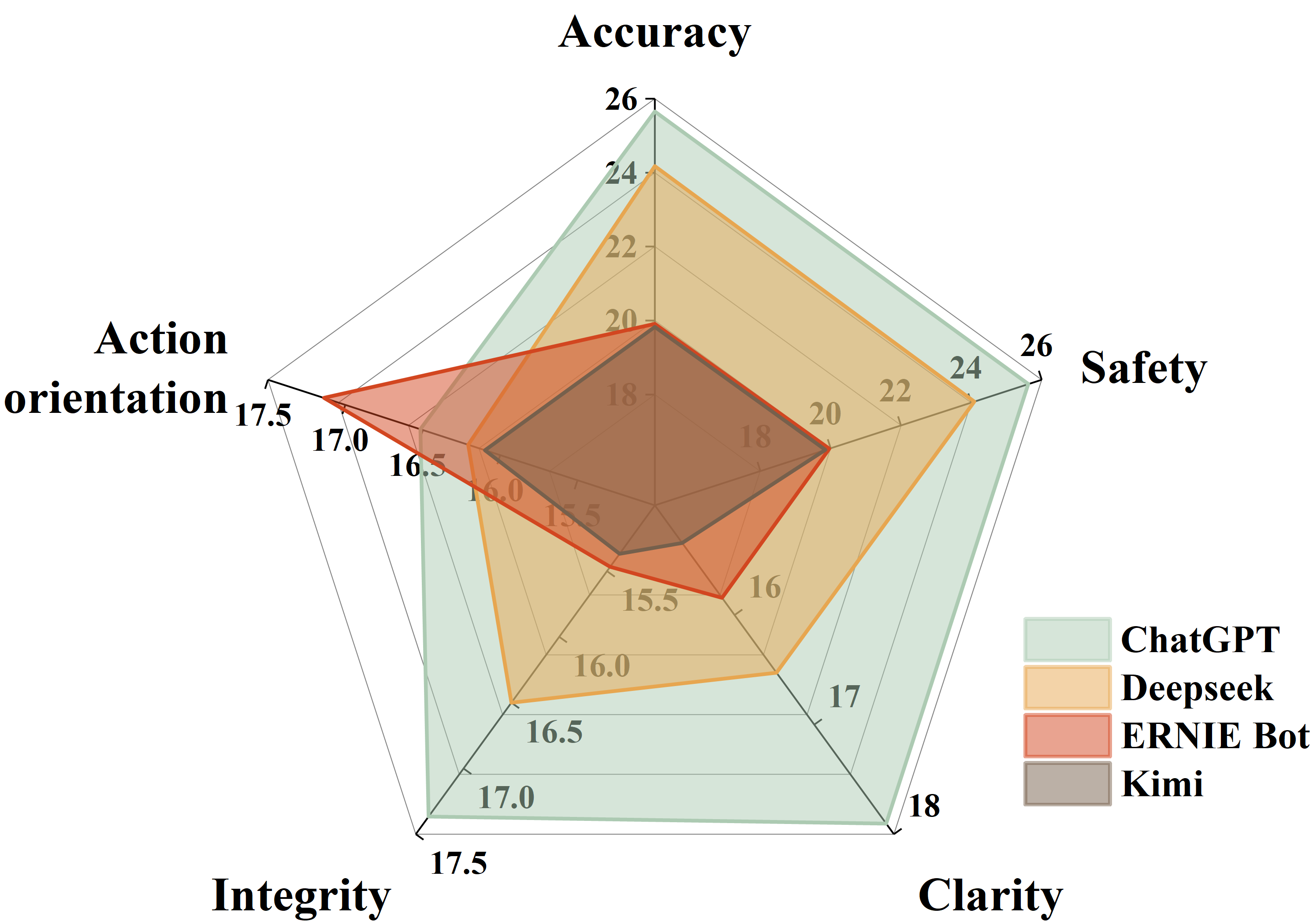}
    \caption{Dimension-wise comparison of AI models across five key quality dimensions:
    Accuracy, Safety, Clarity, Integrity, and Action Orientation. 
    Each polygon represents one AI model’s average score on the five dimensions. 
    Larger enclosed areas indicate stronger overall performance across dimensions.}
    \label{fig:dimension_radar}
\end{figure}

To understand why overall model performance differed, we further examined physician ratings across five evaluation dimensions: \textit{Accuracy}, \textit{Safety}, \textit{Clarity}, \textit{Integrity}, and \textit{Action Orientation} (Figure~\ref{fig:dimension_wise}, Figure~\ref{fig:dimension_radar}). Clear dimension-level disparities emerged across models, revealing distinct strength--weakness profiles.

ChatGPT consistently achieved the highest scores across all five dimensions, including \textit{Accuracy} ($M=25.66$, $SD=2.46$) and \textit{Safety} ($M=25.62$, $SD=2.51$), indicating reliable factual correctness and low-risk phrasing. DeepSeek followed similar patterns but showed noticeable drops in \textit{Clarity} ($M=16.53$) and \textit{Integrity} ($M=16.52$), often producing responses that were accurate yet linguistically fragmented or lacking logical cohesion.

Kimi scored lowest overall, with pronounced weaknesses in \textit{Clarity} ($M=15.53$) and \textit{Integrity} ($M=15.37$), where responses frequently included incomplete reasoning chains or oversimplified advice. ERNIE Bot demonstrated a distinct yet uneven profile: although weaker in factual precision, it performed relatively better in \textit{Action Orientation} ($M=17.14$), tending to generate more procedural and directive recommendations. However, physicians cautioned that such actionable content occasionally lacked appropriate safety framing, highlighting a tension between directness and risk management in patient-facing design.

A two-way repeated-measures ANOVA revealed significant main effects of both Model ($F(3,18)=2315.46$, $p<.001$, $\eta^2=0.713$) and Dimension ($F(4,24)=4823.12$, $p<.001$, $\eta^2=0.922$), as well as a significant interaction ($F(12,72)=351.92$, $p<.001$, $\eta^2=0.606$). These findings confirm that performance differences depend not only on the model but also on the dimension of evaluation, forming distinctive strength–weakness profiles across systems.

Overall, these findings indicate that no single model performs optimally across all quality dimensions. Instead, models exhibit complementary strengths and weaknesses, suggesting the need for adaptive interface strategies that allocate models based on dimension-level reliability. For example, ChatGPT may serve as a default for patient-facing educational content, whereas ERNIE Bot’s procedural strength could be selectively leveraged under additional safety constraints. Such dimension-aware orchestration would allow interactive systems to balance usability and clinical risk more effectively.

\subsection{Qualitative Findings}
\subsubsection{How Physicians Interpret Quality Dimensions in Practice}

Physicians’ qualitative accounts revealed that evaluations of AI-generated health information were grounded in concrete judgment moments rather than abstract notions of quality. In practice, clinicians assessed responses by closely examining how information was framed, prioritized, and translated into guidance, often identifying subtle but consequential breakdowns across multiple quality dimensions. These judgments reflected not only whether content was factually correct, but whether it aligned with clinical expectations for safety, coherence, responsibility boundaries, and appropriate action.

Across interviews, physicians consistently viewed generative AI as a valuable informational scaffold, particularly for supporting patients’ early-stage understanding before clinical encounters. Several participants described AI as a ``pre-visit primer'' that could help patients grasp foundational concepts and formulate more focused questions. As P3 explained, \textit{``As an information aggregator, it performs remarkably well. A newly diagnosed patient might come in already knowing what insulin resistance is, and then ask more specific questions.''} Physicians emphasized that this preparatory role improved the efficiency and depth of doctor--patient communication, especially when explanations were clearly structured and pedagogically framed. Models such as ChatGPT were frequently cited as excelling in this respect, offering coherent narratives and intuitive metaphors that aligned well with clinicians’ expectations for clarity and factual accuracy.

At the same time, several physicians cautioned that linguistic clarity was not a neutral attribute. Rather, fluency and well-organized explanations often amplified the perceived authority of AI outputs, shaping how patients might interpret responsibility and reliability. Some clinicians noted that when clear explanations were not accompanied by explicit boundary-setting, patients could misinterpret educational descriptions as implicit clinical endorsement. In this sense, clarity functioned as a double-edged quality: it supported comprehension, but could also obscure the limits of AI authority if not carefully framed.

However, clinicians were careful to distinguish informational usefulness from clinical appropriateness. A recurring concern centered on \textit{Safety}, particularly in responses that adopted a neutral or balanced tone in situations where explicit prioritization or prohibition was expected. Several physicians described instances in which AI systems presented advantages and disadvantages symmetrically, without clearly signaling when an option should be avoided or when professional consultation was non-negotiable. P6 remarked that \textit{``this kind of neutrality can be dangerous for high-risk patients.''} Notably, physicians observed that such safety failures were rarely due to incorrect facts. Instead, they often stemmed from the omission or de-emphasis of a single critical warning, such as burying disclaimers late in the response or phrasing them too softly to convey urgency. In these cases, otherwise fluent and balanced explanations were perceived as more misleading than explicitly cautious or incomplete ones. ChatGPT was generally regarded as more conservative in risk signaling, whereas other models were more likely to understate safety constraints.

The dimension of \textit{Action Orientation} further exposed tensions in AI-mediated guidance. Physicians differed in their tolerance for directive language, but shared a clear boundary regarding clinical authority. As P2 stated, \textit{``I’m fine with AI explaining options, but once it tells a patient what dosage to change, that crosses a line.''} Some models were criticized for stopping short of actionable advice, frequently deferring responsibility with generic statements such as ``consult your doctor,'' which clinicians felt could frustrate patients seeking practical guidance. Others—most notably ERNIE Bot—were perceived as overly action-oriented, providing step-by-step instructions without sufficient contextual qualification. Physicians emphasized that procedural guidance was only acceptable when tightly coupled with explicit conditions, such as specifying when an action should be taken, when it should be avoided, and what symptoms should trigger escalation to professional care. When these contextual boundaries were absent, actionability itself became a source of clinical risk.

Physicians also evaluated AI responses through the lens of \textit{Integrity}, focusing on internal coherence and the prioritization of information. Several participants criticized outputs that appeared comprehensive yet failed to guide patients toward interpretation or next steps. P3 described this pattern as receiving \textit{``data points, not strategies,''} noting that long lists of possible explanations or recommendations could overwhelm rather than assist patients. DeepSeek was frequently cited as exhibiting this limitation: while often accurate at the sentence level, its responses were described as fragmented or insufficiently synthesized, requiring clinicians to reconstruct a coherent narrative. By contrast, ChatGPT was perceived as more successful at organizing information hierarchically, though physicians cautioned that even well-structured explanations could fall short if they did not adequately foreground risk or responsibility.

In sum, these qualitative findings illustrate how physicians operationalize quality dimensions in real-world evaluation. Rather than treating accuracy, safety, clarity, integrity, and action orientation as independent criteria, clinicians assessed how these dimensions interacted within specific judgment moments. Importantly, physicians often compared models implicitly through these interactions, reasoning about relative strengths and weaknesses rather than evaluating systems in isolation. ChatGPT was viewed as balanced and generally reliable but sometimes conservative; DeepSeek as accurate yet fragmented; ERNIE Bot as highly action-oriented but risky; and Kimi Chat as inconsistent across dimensions. Breakdowns were rarely attributed to incorrect information alone, but instead to failures of contextualization, prioritization, or misalignment with clinical responsibility. These insights help explain the dimension-level trade-offs observed in our quantitative analysis and underscore the importance of aligning AI behavior with how quality is interpreted in practice.

\subsubsection{Implications for Designing Dimension-aware Health AI Systems}

Physicians’ qualitative reflections indicate that many observed limitations of current generative AI systems stem not from insufficient medical knowledge, but from systematic misalignment with how quality dimensions are evaluated in clinical practice. Rather than treating accuracy, safety, clarity, integrity, and action orientation as independent attributes, physicians emphasized that these dimensions must be coordinated through interface-level mediation in order for AI-generated guidance to remain interpretable, safe, and clinically appropriate.

First, physicians stressed that actionable guidance becomes meaningful only when it is grounded in patient-specific context. In our study, breakdowns in \textit{Action Orientation} frequently co-occurred with failures of \textit{Integrity} when systems issued broadly applicable recommendations without accounting for comorbidities, treatment history, or lived constraints. As P6 explained, \textit{``I want AI to stop answering like a textbook. If it could say, `Because this patient has nephropathy, avoid high-protein diets,’ that’s when it becomes truly useful.''} Importantly, physicians did not frame personalization as an optional enhancement, but as a prerequisite for safe action. This suggests that future systems should condition the specificity and assertiveness of recommendations on available patient context and clinical risk, rather than applying a uniform action-oriented style across queries.

Second, concerns about \textit{Safety} were closely tied to the opacity of AI-generated outputs. Physicians repeatedly noted that even factually correct responses became difficult to trust when information sources, uncertainty, or guideline alignment were not made explicit. As P2 remarked, \textit{``I need to know where the information is coming from. Was it a national guideline? A journal article? A user forum?''} These observations indicate that transparency functions as a core safety mechanism rather than a cosmetic interface feature. Making provenance, confidence, or evidence alignment visible enables clinicians and patients alike to interpret AI outputs as situated informational inputs, instead of implicit clinical judgments.

Third, physicians highlighted the importance of contextual grounding for maintaining \textit{Integrity}. Breakdowns frequently occurred when AI systems presented comprehensive but decontextualized information, leaving patients uncertain about relevance or next steps. Several participants described such responses as technically informative yet practically disorienting. To address this, physicians envisioned systems that could anchor explanations in concrete patient data or recent behaviors. For example, P5 suggested that \textit{``If AI could pull my patient’s continuous glucose monitor readings or even step counts, it could explain why their blood sugar spiked last night—and what to do next.''} In this framing, contextual integration was not merely about richer data access, but about restoring narrative coherence—helping patients connect abstract medical knowledge to their lived experience.

Finally, physicians consistently emphasized the need for explicit governance mechanisms to regulate AI autonomy in high-stakes scenarios. Overly directive responses were perceived as particularly risky when they encroached on clinical decision-making without appropriate safeguards. As P7 stated, \textit{``It should be a draft assistant. Let it write, but require a doctor to review before the patient sees anything involving medication changes.''} Such comments point to the importance of dynamically constraining \textit{Action Orientation} based on risk, for instance by gating certain recommendations, flagging outputs for review, or clearly delineating AI-generated drafts from clinician-endorsed guidance.

In conclusion, these implications suggest that effective health AI systems should be designed not as autonomous answer engines, but as dimension-aware mediators that actively align system behavior with the evaluative criteria clinicians apply in practice. By coordinating personalization, transparency, contextual grounding, and governance mechanisms around quality dimensions such as safety, integrity, and action orientation, designers can better balance usefulness, accountability, and trust in AI-assisted health communication.

\section{DISCUSSION}
\subsection{Generative AI as a Scalable Health Educator}

Our findings underscore the value of generative AI as an accessible and scalable tool for patient education, particularly in the management of chronic diseases such as T2DM. Across both quantitative and qualitative analyses, physicians consistently acknowledged the ability of LLMs to deliver structured, accurate, and digestible explanations for factual knowledge and general lifestyle guidance. This aligns with recent work demonstrating that LLMs can support patient understanding of medical terminology, risk factors, and self-care protocols in controlled settings \cite{akrimi2025chatgpt, hernandez2023future}.

What distinguishes our study, however, is the ecological validity of this finding in the context of real-world patient queries and localized healthcare constraints. Unlike prior work that focuses on static prompts or expert-authored question sets \cite{shin2020autoprompt, mou2024sg}, our question corpus emerged directly from patient-initiated information-seeking behavior, collected through observational and participatory methods. This grounding enables a more nuanced understanding of where AI succeeds—as a “pre-visit primer” that prepares patients to ask better-informed questions—and where it fails. Several physicians in our study noted that AI-enhanced literacy improves clinical encounters by enabling patients to articulate symptoms or concepts more precisely, reducing communication overhead.

However, our results also clarify the boundaries of this educational value. Even in its strongest domains, AI does not substitute for clinical reasoning. Physicians emphasized that factual explanations—no matter how clear—are not equivalent to judgment. This echoes concerns raised in prior work about the dangers of misperceived authority when AI-generated outputs are overly fluent or confident \cite{10.1145/3544548.3581251, rajagopal2025generativeaisupportpatients,asgari2025framework}.

Together, these findings support the design of intelligent interfaces that explicitly frame generative AI as an educational tool rather than a diagnostic engine. Future IUI systems could emphasize this distinction through role-based messaging (“AI educator” vs. “clinical decision-maker”), scaffolding features that encourage users to explore questions without assuming correctness, and interface cues that promote follow-up discussions with human professionals. Such systems could position AI as a “health literacy amplifier” embedded in the patient journey, improving not just knowledge access but also downstream communication quality.

\subsection{Dimension-wise Trade-offs Explain Why Model Quality Is Not One-dimensional}

A key implication of our findings is that the quality of AI-generated health information cannot be adequately characterized by a single aggregate score or a simple model ranking. Instead, our quantitative and qualitative evidence jointly suggests that model behavior is shaped by systematic trade-offs across multiple quality dimensions—accuracy, safety, clarity, integrity, and action orientation. This dimension-wise perspective complements prior evaluations that emphasize variability in factual correctness or linguistic fluency across LLMs \cite{hernandez2023future,samimi2025visual}, and aligns with human-centered AI concerns that surface-level usability signals can mask deeper reliability and appropriateness issues in real use contexts \cite{rosbach2025automation,wysocki2023assessing}.

First, our results highlight that high accuracy does not guarantee clinical appropriateness when integrity is compromised. Even when responses contain largely correct medical statements, insufficient prioritization, weak internal coherence, or overly exhaustive enumerations can reduce interpretability and practical value. This echoes broader HCI and health-informatics critiques that “more information” is not necessarily “better support,” particularly for lay users who rely on interfaces to structure and contextualize guidance \cite{merry2021mental,schrills2020answer}. In our study, integrity emerged as an important mediator between correctness and usefulness: responses that failed to impose hierarchy or translate content into a coherent explanatory narrative were more likely to be perceived as fragmented, even when not strictly incorrect.

Second, safety failures in our dataset were not primarily driven by blatantly erroneous statements, but by subtle breakdowns in risk signaling—such as missing, softened, or de-emphasized warnings. This observation extends prior work on trust calibration and uncertainty communication, which emphasizes that users often interpret fluency and completeness as proxies for reliability \cite{reyes2025trusting,reis2024influence}. In consumer-facing health contexts, such miscalibration is especially consequential because users may act on advice without professional oversight. Our findings therefore support the view that safety should be treated as an interface-level property—achieved through explicit framing, prioritization, and boundary-setting—rather than a passive byproduct of accuracy alone \cite{bhatt2021uncertainty}.

Third, action orientation introduces a fundamental tension between practical usefulness and clinical risk. Actionable recommendations can improve user experience by reducing ambiguity and supporting self-care routines, but they can also amplify harm when the same directive tone is applied to situations that require individualized clinical judgment. Prior work has noted that the persuasive form of AI outputs can shape user reliance even when underlying validity is uncertain \cite{reis2024influence,rosbach2025automation}. Our results reinforce that actionability cannot be treated as uniformly desirable. Instead, it functions as a conditional dimension whose appropriateness depends on whether the system can specify contextual boundaries, escalation triggers, and non-negotiable consultation points.

Finally, model comparisons in our study indicate that these dimension-level trade-offs are not distributed uniformly across systems. Different models may achieve similar aggregate performance while exhibiting qualitatively distinct profiles—for example, stronger fluency but weaker coherence, or higher directive actionability but weaker safety signaling. This aligns with recent arguments that evaluation should move beyond “best model” framing toward more granular, task- and risk-sensitive assessment of model behavior \cite{asgari2025framework,gaber2025evaluating,jahan2024comprehensive}. Taken together, our findings motivate a shift from monolithic notions of model quality toward dimension-aware interpretations, which better reflect how clinicians and patients experience AI outputs in practice.

By foregrounding trade-offs among accuracy, safety, clarity, integrity, and action orientation, this discussion reframes model “quality” as an inherently multi-dimensional construct. This framing provides a stronger foundation for downstream design decisions: rather than selecting a single model based on aggregate scores, health AI systems should consider how dimension-level strengths and weaknesses align with the demands and risks of specific interaction contexts.

\subsection{Calibrated Trust through Dimension-aware Mediation}

A central theoretical implication of our findings is that trust in AI-generated health information cannot be treated as a direct function of output fluency or overall performance. Instead, trust emerges from how users—implicitly or explicitly—interpret multiple quality dimensions in relation to perceived risk, context, and responsibility. Our results therefore challenge a common assumption in the deployment of conversational AI: that improving linguistic naturalness or factual coverage will naturally yield appropriate trust.

Prior work in human–AI interaction has repeatedly shown that fluent and well-structured language can inflate perceived credibility, even when underlying reasoning is incomplete or uncertain \cite{wysocki2023assessing,reis2024influence,rosbach2025automation}. In health contexts, this “fluency-driven trust” is particularly problematic, as users may equate confident phrasing with clinical validity. Our study extends this literature by demonstrating that physicians do not evaluate trustworthiness holistically. Instead, they attend to specific breakdowns across accuracy, safety, integrity, and action orientation—often discounting otherwise fluent responses when these dimensions are misaligned.

Quantitatively, we observed that models with strong aggregate scores still exhibited uneven performance across dimensions, while qualitatively, physicians articulated concrete concerns about how these imbalances could mislead patients. Together, these findings suggest that trust is not merely about whether an answer is correct or readable, but whether its dimension-level signals are appropriately calibrated to the stakes of the interaction. For example, high action orientation without explicit safety framing was perceived as more dangerous than vague but cautious responses, whereas high clarity without prioritization undermined interpretability and decision confidence.

This perspective aligns with prior work on calibrated trust, which argues that effective human–AI systems should support neither blind reliance nor blanket skepticism, but context-sensitive reliance grounded in transparent system behavior \cite{reyes2025trusting,bhatt2021uncertainty}. However, existing approaches often operationalize calibration through uncertainty visualization or confidence scores alone. Our findings suggest that such mechanisms are insufficient if they do not reflect how different quality dimensions interact. A system may accurately communicate uncertainty while still failing to signal responsibility boundaries or appropriate action constraints.

From an IUI standpoint, these results motivate a shift from output-centric trust mechanisms to mediation-centric ones. Rather than treating trust as a property of the model, trust should be shaped at the interface level through dimension-aware mediation—where the system actively modulates presentation, framing, and interaction patterns based on the dimension-level reliability of a response. This includes adjusting tone, highlighting safety boundaries, restructuring content to emphasize prioritization, or suppressing action-oriented phrasing when contextual judgment is required.

Such mediation reframes trust not as a static belief in AI competence, but as an interactional outcome co-produced by system behavior and interface design. In this framing, fluency becomes a necessary but insufficient condition for trust. Instead, appropriate trust depends on whether the system’s signals across accuracy, safety, clarity, integrity, and action orientation are aligned with user expectations and situational risk. By foregrounding dimension-aware mediation, our work contributes to a more nuanced theory of trust in intelligent user interfaces—one that moves beyond surface-level usability toward responsible, context-sensitive reliance in high-stakes domains.

\subsection{Design Implications for Dimension-aware Health IUIs}

Our findings suggest that responsible deployment of generative AI in chronic care requires IUIs to mediate model outputs through dimension-aware interaction mechanisms, rather than presenting responses as uniformly reliable answers. Below we outline design implications that translate the dimension-level trade-offs (Section~6.2) and fluency-driven trust risks (Section~6.3) into interface-level strategies grounded in prior IUI and health-informatics work.

First, IUIs should \textit{surface dimension-level signals} to support calibrated trust, making it easier for users to distinguish between responses that are clear yet non-actionable, actionable yet weakly safety-framed, or comprehensive yet lacking prioritization. Prior work on uncertainty expression and trust calibration has shown that interface cues can shape reliance decisions \cite{bhatt2021uncertainty,reyes2025trusting}. Our results extend this line of work by indicating that uncertainty cues alone are insufficient unless they map onto the specific quality dimensions clinicians attend to (e.g., integrity and action orientation), suggesting a need for richer, multi-dimensional presentation of reliability signals.

Second, IUIs should \textit{mediate actionability through explicit responsibility boundaries}. Because action-oriented language can both improve usability and increase clinical risk, interfaces should modulate how directive a response appears and when escalation is triggered. This aligns with safety-oriented interaction design patterns that emphasize graceful fallback, guarded autonomy, and escalation pathways in high-stakes systems \cite{florins2004graceful}. Concretely, systems can frame outputs as educational drafts rather than recommendations, couple actionable steps with clear “consult clinician” gates, and suppress prescriptive phrasing when dimension-level reliability (e.g., safety or integrity) is weak.

Third, our results motivate \textit{dimension-aware model orchestration}. Since no single model consistently excels across all dimensions, IUIs can act as mediators that dynamically route, constrain, or combine models based on the dimension profile needed for the interaction, rather than selecting a single “best” model globally. This implication builds on emerging work in orchestration and multi-model system design \cite{naik2025designing,lee2025map}, while our study provides clinician-grounded evidence for why orchestration should be driven by quality dimensions (accuracy, safety, clarity, integrity, action orientation) rather than by output fluency or aggregate scores alone.

Fourth, several physician expectations point to the importance of \textit{dynamic personalization} as a mechanism for improving both integrity and action orientation. Participants highlighted that generic recommendations often fail not because they are incorrect, but because they are insufficiently contextualized to comorbidities, routines, and constraints—suggesting that future IUIs should incorporate patient context when available and ethically permissible. This direction is consistent with prior work on personalized health interfaces and behaviorally informed recommendations \cite{ayobi2021co,torkamaan2022recommendations,jacobs2018mypath}, and suggests opportunities to integrate structured patient data or self-tracking signals to scaffold context-sensitive guidance.

Finally, IUIs should treat \textit{emotional responsiveness} as part of interface mediation rather than surface-level wording. Our qualitative results suggest that fluency can create an illusion of empathy, while users still perceive emotional support as generic or misattuned. This resonates with critiques of “empathetic language” that lacks interactional grounding and may distort trust \cite{gencc2024situating,cuadra2024illusion}. For chronic care contexts, emotion-aware mediation can be operationalized through pacing, validation, and supportive resource linking, while maintaining explicit boundaries around clinical authority.

Taken together, these implications suggest a shift from answer-centric AI toward dimension-aware mediation in health IUIs. By grounding model outputs in calibrated trust cues, responsibility boundaries, orchestration, personalization, and emotionally attuned interaction, interface design becomes the primary lever for aligning generative AI assistance with clinical expectations and real-world safety needs.

\subsection{Limitations and Future Work}

This study has several limitations that contextualize the scope of our findings and point to directions for future research. First, although we evaluated four widely used generative AI models, our results capture model behavior at a specific moment in time. Given the rapid pace of model iteration and safety alignment, the relative performance and dimension-level trade-offs observed here may evolve. Longitudinal evaluations will be necessary to examine whether improvements in fluency, alignment, or training data translate into more balanced performance across accuracy, safety, integrity, and action orientation.

Second, our evaluation focused on single-turn responses, which limits our ability to assess how quality dimensions unfold in interactive settings. Real-world health information seeking often involves multi-turn clarification, follow-up questions, and user corrections. Future work should extend dimension-aware evaluation to interactive scenarios, examining how trust calibration, actionability, and safety signaling change over the course of an interaction. Such studies would also allow investigation of whether interface-level mediation strategies—such as fallback prompts or escalation cues—can dynamically correct early dimension-level breakdowns.

Third, our qualitative insights were derived from interviews with seven endocrinologists affiliated with urban hospitals in China. While this sample provided clinically grounded perspectives on diabetes care, it may not reflect the expectations or constraints of other healthcare roles or settings, such as primary care, rural practice, nursing, or patient self-management without regular clinical supervision. Expanding participant diversity will be important for understanding how dimension-level quality and trust calibration are interpreted across different clinical workflows and cultural contexts.

Finally, while prompts were standardized across models, we did not systematically vary prompt structure or framing. Prior work has shown that prompt phrasing can influence both model behavior and user perception. Future research should investigate how prompt design interacts with quality dimensions—for example, whether explicit boundary-setting prompts can improve safety signaling or whether certain framings exacerbate fluency-driven overreliance. Together, these directions highlight the need for continued investigation into how dimension-aware evaluation and interface mediation can be generalized, validated, and operationalized in real-world health IUIs.

\section{CONCLUSION}
This paper examined how contemporary generative AI systems support T2DM self-management, drawing on formative patient grounding and a dimension-based evaluation by physicians. Our findings reveal persistent tensions at the dimension level—between accessibility and appropriateness, fluency and clinical reliability, and standardized responses and context-sensitive clinical judgment—highlighting how seemingly high-quality AI outputs can still break down in chronic care settings.

By treating evaluation dimensions as an interpretive lens rather than a fixed framework, this work foregrounds the role of intelligent user interfaces in mediating trust, responsibility, and action in AI-assisted health communication. We argue that long-term conditions such as T2DM call for IUI designs that explicitly support calibrated trust, contextual awareness, and responsible boundary-setting, rather than relying on fluent generation alone.

\bibliographystyle{ACM-Reference-Format}
\bibliography{main}

@article{braun2006using,
  title={Using thematic analysis in psychology},
  author={Braun, Virginia and Clarke, Victoria},
  journal={Qualitative research in psychology},
  volume={3},
  number={2},
  pages={77--101},
  year={2006},
  publisher={Taylor \& Francis}
}

@online{openai_chatgpt_2025,
  author       = {OpenAI},
  title        = {{ChatGPT} (June 2025 Version)},
  year         = {2025},
  url          = {https://openai.com/},
  note         = {OpenAI ChatGPT large language model, accessed June 2025},
  urldate      = {2025-06-01}
}

@online{deepseek_2025,
  author       = {DeepSeek},
  title        = {{DeepSeek} (June 2025 Version)},
  year         = {2025},
  url          = {https://www.deepseek.com/},
  note         = {DeepSeek large language model, accessed June 2025},
  urldate      = {2025-06-01}
}

@online{kimi_2025,
  author       = {Moonshot AI},
  title        = {{Kimi} (June 2025 Version)},
  year         = {2025},
  url          = {https://www.kimi.com/},
  note         = {Kimi AI assistant developed by Moonshot AI, accessed June 2025},
  urldate      = {2025-06-01}
}

@online{erniebot_2025,
  author       = {Baidu},
  title        = {{ERNIE Bot} (June 2025 Version)},
  year         = {2025},
  url          = {https://yiyan.baidu.com/},
  note         = {ERNIE Bot developed by Baidu, accessed June 2025},
  urldate      = {2025-06-01}
}

@inproceedings{10.1145/3544548.3581251,
author = {Burgess, Eleanor R. and Jankovic, Ivana and Austin, Melissa and Cai, Nancy and Kapu\'{s}ci\'{n}ska, Adela and Currie, Suzanne and Overhage, J. Marc and Poole, Erika S and Kaye, Jofish},
title = {Healthcare AI Treatment Decision Support: Design Principles to Enhance Clinician Adoption and Trust},
year = {2023},
isbn = {9781450394215},
publisher = {Association for Computing Machinery},
address = {New York, NY, USA},
url = {https://doi.org/10.1145/3544548.3581251},
doi = {10.1145/3544548.3581251},
booktitle = {Proceedings of the 2023 CHI Conference on Human Factors in Computing Systems},
articleno = {15},
numpages = {19},
location = {Hamburg, Germany},
series = {CHI '23}
}

@inproceedings{10.1145/3290605.3300600,
author = {Desai, Pooja M. and Mitchell, Elliot G. and Hwang, Maria L. and Levine, Matthew E. and Albers, David J. and Mamykina, Lena},
title = {Personal Health Oracle: Explorations of Personalized Predictions in Diabetes Self-Management},
year = {2019},
isbn = {9781450359702},
publisher = {Association for Computing Machinery},
address = {New York, NY, USA},
url = {https://doi.org/10.1145/3290605.3300600},
doi = {10.1145/3290605.3300600},
booktitle = {Proceedings of the 2019 CHI Conference on Human Factors in Computing Systems},
pages = {1–13},
numpages = {13},
location = {Glasgow, Scotland Uk},
series = {CHI '19}
}

@inproceedings{samimi2025visual,
  title={Visual-Conversational Interface for Evidence-Based Explanation of Diabetes Risk Prediction},
  author={Samimi, Reza and Bhattacharya, Aditya and Gosak, Lucija and Stiglic, Gregor and Verbert, Katrien},
  booktitle={Proceedings of the 7th ACM Conference on Conversational User Interfaces},
  pages={1--18},
  year={2025}
}

@article{hernandez2023future,
  title={The future of patient education: AI-driven guide for type 2 diabetes},
  author={Hernandez, Carlos A and Gonzalez, Andres E Vazquez and Polianovskaia, Anastasiia and Sanchez, Rafael Amoro and Arce, Veronica Muyolema and Mustafa, Ahmed and Vypritskaya, Ekaterina and Gutierrez, Oscar Perez and Bashir, Muhammad and Sedeh, Ashkan Eighaei},
  journal={Cureus},
  volume={15},
  number={11},
  year={2023},
  publisher={Cureus}
}

@article{akrimi2025chatgpt,
  title={ChatGPT-4o-Generated Exercise Plans for Patients with Type 2 Diabetes Mellitus—Assessment of Their Safety and Other Quality Criteria by Coaching Experts},
  author={Akrimi, Samir and Schwensfeier, Leon and D{\"u}king, Peter and Kreutz, Thorsten and Brinkmann, Christian},
  journal={Sports},
  volume={13},
  number={4},
  pages={92},
  year={2025},
  publisher={MDPI}
}

@article{gong2020my,
  title={My diabetes coach, a mobile app--based interactive conversational agent to support type 2 diabetes self-management: randomized effectiveness-implementation trial},
  author={Gong, Enying and Baptista, Shaira and Russell, Anthony and Scuffham, Paul and Riddell, Michaela and Speight, Jane and Bird, Dominique and Williams, Emily and Lotfaliany, Mojtaba and Oldenburg, Brian},
  journal={Journal of medical Internet research},
  volume={22},
  number={11},
  pages={e20322},
  year={2020},
  publisher={JMIR Publications Toronto, Canada}
}

@article{lanshan2025factors,
  title={Factors influencing the acceptance of medical AI chat assistants among healthcare professionals and patients: a survey-based study in China},
  author={Lanshan, Zhang and Yang, Jingyi and Fang, Gege},
  journal={Frontiers in Public Health},
  volume={13},
  pages={1637270},
  year={2025},
  publisher={Frontiers}
}

@article{guo2025promoting,
  title={Promoting trust and intention to adopt health information generated by ChatGPT among healthcare customers: An empirical study},
  author={Guo, Shuangyan and Song, Yang and Chen, Guanyun and Han, Hongxin and Wu, Hong and Ma, Jingdong},
  journal={Digital Health},
  volume={11},
  pages={20552076251374121},
  year={2025},
  publisher={SAGE Publications Sage UK: London, England}
}

@misc{rajagopal2025generativeaisupportpatients,
    title={Can Generative AI Support Patients' \& Caregivers' Informational Needs? Towards Task-Centric Evaluation Of AI Systems}, 
    author={Shreya Rajagopal and Jae Ho Sohn and Hari Subramonyam and Shiwali Mohan},
    year={2025},
    eprint={2402.00234},
    archivePrefix={arXiv},
    primaryClass={cs.HC},
    url={https://arxiv.org/abs/2402.00234}, 
}

@article{mirbabaie2025digital,
  title={Digital Assistants for Diabetes Treatment: Designing a User Interface to Support Chronic Disease Self-Management},
  author={Mirbabaie, Milad and Marx, Julian and M{\"o}llmann, Nicholas and Matt, Svenja},
  journal={ACM SIGMIS Database: the DATABASE for Advances in Information Systems},
  volume={56},
  number={3},
  pages={55--79},
  year={2025},
  publisher={ACM New York, NY, USA}
}

@article{mitchell2021automated,
  title={Automated vs. human health coaching: exploring participant and practitioner experiences},
  author={Mitchell, Elliot G and Maimone, Rosa and Cassells, Andrea and Tobin, Jonathan N and Davidson, Patricia and Smaldone, Arlene M and Mamykina, Lena},
  journal={Proceedings of the ACM on human-computer interaction},
  volume={5},
  number={CSCW1},
  pages={1--37},
  year={2021},
  publisher={ACM New York, NY, USA}
}

@inproceedings{gerstenberg2025living,
  title={Living Well with Diabetes: Rethinking Digital Diabetes Management Systems},
  author={Gerstenberg, Rebecca and Neuhaus, Robin and Vitt, Martin and Hassenzahl, Marc},
  booktitle={Proceedings of the 2025 ACM Designing Interactive Systems Conference},
  pages={3153--3172},
  year={2025}
}

@article{mayberry2019mhealth,
  title={mHealth interventions for disadvantaged and vulnerable people with type 2 diabetes},
  author={Mayberry, Lindsay Satterwhite and Lyles, Courtney R and Oldenburg, Brian and Osborn, Chandra Y and Parks, Makenzie and Peek, Monica E},
  journal={Current diabetes reports},
  volume={19},
  number={12},
  pages={148},
  year={2019},
  publisher={Springer}
}

@article{dsouza2024identification,
  title={Identification of challenges and leveraging mHealth technology, with need-based solutions to empower self-management in type 2 diabetes: a qualitative study},
  author={Dsouza, Sherize Merlin and Venne, Julien and Shetty, Sahana and Brand, Helmut},
  journal={Diabetology \& Metabolic Syndrome},
  volume={16},
  number={1},
  pages={182},
  year={2024},
  publisher={Springer}
}

@article{biernatzki2018information,
  title={Information needs in people with diabetes mellitus: a systematic review},
  author={Biernatzki, Lisa and Kuske, Silke and Genz, Jutta and Ritschel, Michaela and Stephan, Astrid and B{\"a}chle, Christina and Droste, Sigrid and Grobosch, Sandra and Ernstmann, Nicole and Chernyak, Nadja and others},
  journal={Systematic reviews},
  volume={7},
  number={1},
  pages={27},
  year={2018},
  publisher={Springer}
}

@inproceedings{bhattacharya2023directive,
  title={Directive explanations for monitoring the risk of diabetes onset: introducing directive data-centric explanations and combinations to support what-if explorations},
  author={Bhattacharya, Aditya and Ooge, Jeroen and Stiglic, Gregor and Verbert, Katrien},
  booktitle={Proceedings of the 28th international conference on intelligent user interfaces},
  pages={204--219},
  year={2023}
}

@article{ayobi2021co,
  title={Co-designing personal health? Multidisciplinary benefits and challenges in informing diabetes self-care technologies},
  author={Ayobi, Amid and Stawarz, Katarzyna and Katz, Dmitri and Marshall, Paul and Yamagata, Taku and Santos-Rodriguez, Raul and Flach, Peter and O'Kane, Aisling Ann},
  journal={Proceedings of the ACM on Human-Computer Interaction},
  volume={5},
  number={CSCW2},
  pages={1--26},
  year={2021},
  publisher={ACM New York, NY, USA}
}

@article{rebitschek2025evaluating,
  title={Evaluating evidence-based health information from generative AI using a cross-sectional study with laypeople seeking screening information},
  author={Rebitschek, Felix G and Carella, Alessandra and Kohlrausch-Pazin, Silja and Zitzmann, Michael and Steckelberg, Anke and Wilhelm, Christoph},
  journal={npj Digital Medicine},
  volume={8},
  number={1},
  pages={343},
  year={2025},
  publisher={Nature Publishing Group UK London}
}

@article{reddy2024generative,
  title={Generative AI in healthcare: an implementation science informed translational path on application, integration and governance},
  author={Reddy, Sandeep},
  journal={Implementation Science},
  volume={19},
  number={1},
  pages={27},
  year={2024},
  publisher={Springer}
}

@article{agarwal2024medhalu,
  title={Medhalu: Hallucinations in responses to healthcare queries by large language models},
  author={Agarwal, Vibhor and Jin, Yiqiao and Chandra, Mohit and De Choudhury, Munmun and Kumar, Srijan and Sastry, Nishanth},
  journal={arXiv preprint arXiv:2409.19492},
  year={2024}
}

@article{jin2019pubmedqa,
  title={Pubmedqa: A dataset for biomedical research question answering},
  author={Jin, Qiao and Dhingra, Bhuwan and Liu, Zhengping and Cohen, William W and Lu, Xinghua},
  journal={arXiv preprint arXiv:1909.06146},
  year={2019}
}

@article{kim2024medexqa,
  title={MedExQA: Medical question answering benchmark with multiple explanations},
  author={Kim, Yunsoo and Wu, Jinge and Abdulle, Yusuf and Wu, Honghan},
  journal={arXiv preprint arXiv:2406.06331},
  year={2024}
}

@article{kim2024human,
  title={Human-centered evaluation of explainable AI applications: a systematic review},
  author={Kim, Jenia and Maathuis, Henry and Sent, Danielle},
  journal={Frontiers in Artificial Intelligence},
  volume={7},
  pages={1456486},
  year={2024},
  publisher={Frontiers Media SA}
}

@inproceedings{cuadra2024illusion,
  title={The illusion of empathy? notes on displays of emotion in human-computer interaction},
  author={Cuadra, Andrea and Wang, Maria and Stein, Lynn Andrea and Jung, Malte F and Dell, Nicola and Estrin, Deborah and Landay, James A},
  booktitle={Proceedings of the 2024 CHI Conference on Human Factors in Computing Systems},
  pages={1--18},
  year={2024}
}

@article{swallow2025digibete,
  title={DigiBete, a novel chatbot to support transition to adult care of young people/young adults with type 1 diabetes mellitus: outcomes from a prospective, multimethod, nonrandomized feasibility and acceptability study},
  author={Swallow, Veronica and Horsman, Janet and Mazlan, Eliza and Campbell, Fiona and Zaidi, Reza and Julian, Madeleine and Branchflower, Jacob and Martin-Kerry, Jackie and Monks, Helen and Soni, Astha and others},
  journal={JMIR diabetes},
  volume={10},
  number={1},
  pages={e74032},
  year={2025},
  publisher={JMIR Publications Inc., Toronto, Canada}
}

@inproceedings{tseng2025designing,
  title={Designing an AI Chatbot for Team-Based Diabetes Care: An Iterative Human-in-the-Loop Approach},
  author={Tseng, Yuan-Chi and Chen, Samuel and Mah, Kang-Heng and Chen, Yu-Chang},
  booktitle={International Conference on Human-Computer Interaction},
  pages={261--276},
  year={2025},
  organization={Springer}
}

@inproceedings{schombs2025designing,
  title={Designing Embodied Agents for Impactful Human-Centred Risk Communication},
  author={Sch{\"o}mbs, Sarah},
  booktitle={Proceedings of the Extended Abstracts of the CHI Conference on Human Factors in Computing Systems},
  pages={1--5},
  year={2025}
}

@article{shin2020autoprompt,
  title={Autoprompt: Eliciting knowledge from language models with automatically generated prompts},
  author={Shin, Taylor and Razeghi, Yasaman and Logan IV, Robert L and Wallace, Eric and Singh, Sameer},
  journal={arXiv preprint arXiv:2010.15980},
  year={2020}
}

@article{mou2024sg,
  title={Sg-bench: Evaluating llm safety generalization across diverse tasks and prompt types},
  author={Mou, Yutao and Zhang, Shikun and Ye, Wei},
  journal={Advances in Neural Information Processing Systems},
  volume={37},
  pages={123032--123054},
  year={2024}
}

@article{asgari2025framework,
  title={A framework to assess clinical safety and hallucination rates of LLMs for medical text summarisation},
  author={Asgari, Elham and Monta{\~n}a-Brown, Nina and Dubois, Magda and Khalil, Saleh and Balloch, Jasmine and Yeung, Joshua Au and Pimenta, Dominic},
  journal={npj Digital Medicine},
  volume={8},
  number={1},
  pages={274},
  year={2025},
  publisher={Nature Publishing Group UK London}
}

@article{gaber2025evaluating,
  title={Evaluating large language model workflows in clinical decision support for triage and referral and diagnosis},
  author={Gaber, Farieda and Shaik, Maqsood and Allega, Fabio and Bilecz, Agnes Julia and Busch, Felix and Goon, Kelsey and Franke, Vedran and Akalin, Altuna},
  journal={npj Digital Medicine},
  volume={8},
  number={1},
  pages={263},
  year={2025},
  publisher={Nature Publishing Group UK London}
}

@article{jahan2024comprehensive,
  title={A comprehensive evaluation of large language models on benchmark biomedical text processing tasks},
  author={Jahan, Israt and Laskar, Md Tahmid Rahman and Peng, Chun and Huang, Jimmy Xiangji},
  journal={Computers in biology and medicine},
  volume={171},
  pages={108189},
  year={2024},
  publisher={Elsevier}
}

@article{reis2024influence,
  title={Influence of believed AI involvement on the perception of digital medical advice},
  author={Reis, Moritz and Reis, Florian and Kunde, Wilfried},
  journal={Nature Medicine},
  volume={30},
  number={11},
  pages={3098--3100},
  year={2024},
  publisher={Nature Publishing Group US New York}
}

@inproceedings{rosbach2025automation,
  title={Automation Bias in AI-assisted Medical Decision-making under Time Pressure in Computational Pathology},
  author={Rosbach, Emely and Ganz, Jonathan and Ammeling, Jonas and Riener, Andreas and Aubreville, Marc},
  booktitle={BVM Workshop},
  pages={129--134},
  year={2025},
  organization={Springer}
}

@article{wysocki2023assessing,
  title={Assessing the communication gap between AI models and healthcare professionals: Explainability, utility and trust in AI-driven clinical decision-making},
  author={Wysocki, Oskar and Davies, Jessica Katharine and Vigo, Markel and Armstrong, Anne Caroline and Landers, D{\'o}nal and Lee, Rebecca and Freitas, Andr{\'e}},
  journal={Artificial Intelligence},
  volume={316},
  pages={103839},
  year={2023},
  publisher={Elsevier}
}

@article{reyes2025trusting,
  title={Trusting AI: does uncertainty visualization affect decision-making?},
  author={Reyes, Jonatan and Batmaz, Anil Ufuk and Kersten-Oertel, Marta},
  journal={Frontiers in Computer Science},
  volume={7},
  pages={1464348},
  year={2025},
  publisher={Frontiers Media SA}
}

@inproceedings{bhatt2021uncertainty,
  title={Uncertainty as a form of transparency: Measuring, communicating, and using uncertainty},
  author={Bhatt, Umang and Antor{\'a}n, Javier and Zhang, Yunfeng and Liao, Q Vera and Sattigeri, Prasanna and Fogliato, Riccardo and Melan{\c{c}}on, Gabrielle and Krishnan, Ranganath and Stanley, Jason and Tickoo, Omesh and others},
  booktitle={Proceedings of the 2021 AAAI/ACM Conference on AI, Ethics, and Society},
  pages={401--413},
  year={2021}
}

@article{merry2021mental,
  title={A mental models approach for defining explainable artificial intelligence},
  author={Merry, Michael and Riddle, Pat and Warren, Jim},
  journal={BMC Medical Informatics and Decision Making},
  volume={21},
  number={1},
  pages={344},
  year={2021},
  publisher={Springer}
}

@article{schrills2020answer,
  title={How to Answer Why--Evaluating the Explanations of AI Through Mental Model Analysis},
  author={Schrills, Tim and Franke, Thomas},
  journal={arXiv preprint arXiv:2002.02526},
  year={2020}
}

@inproceedings{torkamaan2022recommendations,
  title={Recommendations as challenges: estimating required effort and user ability for health behavior change recommendations},
  author={Torkamaan, Helma and Ziegler, J{\"u}rgen},
  booktitle={Proceedings of the 27th International Conference on Intelligent User Interfaces},
  pages={106--119},
  year={2022}
}

@article{jacobs2018mypath,
  title={MyPath: Investigating breast cancer patients' use of personalized health information},
  author={Jacobs, Maia and Johnson, Jeremy and Mynatt, Elizabeth D},
  journal={Proceedings of the ACM on Human-Computer Interaction},
  volume={2},
  number={CSCW},
  pages={1--21},
  year={2018},
  publisher={ACM New York, NY, USA}
}

@inproceedings{naik2025designing,
  title={Designing with Multi-Agent Generative AI: Insights from Industry Early Adopters},
  author={Naik, Suchismita and Toombs, Austin L and Snellinger, Amanda and Saponas, Scott and Hall, Amanda K},
  booktitle={Proceedings of the 2025 ACM Designing Interactive Systems Conference},
  pages={1961--1972},
  year={2025}
}

@inproceedings{lee2025map,
  title={MAP: Multi-user Personalization with Collaborative LLM-powered Agents},
  author={Lee, Christine P and Choi, Jihye and Mutlu, Bilge},
  booktitle={Proceedings of the Extended Abstracts of the CHI Conference on Human Factors in Computing Systems},
  pages={1--11},
  year={2025}
}

@inproceedings{florins2004graceful,
  title={Graceful degradation of user interfaces as a design method for multiplatform systems},
  author={Florins, Murielle and Vanderdonckt, Jean},
  booktitle={Proceedings of the 9th international conference on Intelligent user interfaces},
  pages={140--147},
  year={2004}
}

@article{gencc2024situating,
  title={Situating Empathy in HCI/CSCW: A Scoping Review},
  author={Gen{\c{c}}, U{\u{g}}ur and Verma, Himanshu},
  journal={Proceedings of the ACM on Human-Computer Interaction},
  volume={8},
  number={CSCW2},
  pages={1--37},
  year={2024},
  publisher={ACM New York, NY, USA}
}
\end{document}